\documentclass[10pt]{article}
\usepackage{amsmath}
\usepackage{amssymb}
\usepackage{color,tikz,caption}
\usetikzlibrary{decorations.markings,arrows, calc}
\def \nn{\nonumber}
\usepackage{wrapfig}
\def\bea#1\eea{\begin{align}#1\end{align}}

\usepackage{soul}
\usepackage{framed}
\usepackage{amssymb,cancel}
\usepackage{wrapfig}
\usepackage{mathrsfs}
\usepackage{amsbsy}
\usepackage{color}
\usepackage{graphicx}
\usepackage[colorlinks=true, pdfstartview=FitV, linkcolor=cyan, citecolor=blue, urlcolor=blue]{hyperref}
\definecolor{shadecolor}{rgb}{0.95, 0.95, 0.86}

\def\P{\mathbf P}

\def\replace#1#2{\red{\cancel{{\hbox{\scriptsize  {$#1$}} }}}{\blue {#2}}}

\def\red#1{\textcolor[rgb]{0.9, 0, 0}{#1} }

\def\blue#1{\textcolor[rgb]{0,0,1}{#1}}

\usepackage{verbatim}
\def\wt{\widetilde}

\def\res{\mathop{\mathrm{res}}\limits_}

\renewcommand{\theequation}{\arabic{section}-\arabic{equation}}
\makeatletter
\@addtoreset{equation}{section}
\makeatother
\def\tr{\mathrm {Tr}}
\def\le{\left}
\def \div{ {\rm div}}
\def\ri{\right}
\def\QED{\hfill $\blacksquare$\par\vskip 5pt}

\def\bc{\begin{corollary}}
\def\ec{\end{corollary}}

\def \s{\mathfrak s}

\def\br{\begin{remark}\rm\small}
\def\1{{\bf 1}}
\def\er{\end{remark}}
\def\bt{
\begin{shaded}
\begin{theorem}}
\def\et{\end{theorem}
\end{shaded}}
\def\bd{
\begin{shaded}
\begin{definition}}
\def\ed{\end{definition}
\end{shaded}}
\def\bp{
\begin{shaded}
\begin{proposition}}
\def\ep{\end{proposition}
\end{shaded}}

\def\bx{\begin{example}}
\def\ex{\end{example}}
\def\bi{\begin{itemize}}
\def\ei{\end{itemize}}
\def\bd{\begin{definition}}
\def\ed{\end{definition}}
\def\bp{\begin{proposition}\rm}
\def\bl{\begin{lemma}\em}
\def\el{\end{lemma}}
\def\ep{\end{proposition}}
\def \pa{\partial}
\def\C{{\mathbb C}}
\def\R{{\mathbb R}}
\def\N{{\mathbb N}}
\def\Z{{\mathbb Z}}

\newtheorem{theorem}{Theorem}[section]
\newtheorem{example}{Example}[section]
\newtheorem{corollary}{Corollary}[section]

\newtheorem{examps}{Examples}[section]

\newtheorem{lemma}{Lemma}[section]
\newtheorem{remark}{Remark}[section]
\newtheorem{remarks}[remark]{Remarks}
\newtheorem{proposition}{Proposition}[section] 
\newtheorem{definition}{Definition}[section]
\def\br{\begin{remark}}
\def\er{\end{remark}}
\def\bl{\begin{lemma}}
\def\el{\end{lemma}}
\def\bxs{\begin{examps}. \rm\begin{enumerate}}
\def\exs{\end{enumerate}\end{examps}}
\def\bd{\begin{definition}}
\def\ed{\end{definition}}
\def\bp{\begin{proposition}}
\def\ep{\end{proposition}}
\def\be{\begin{equation}}
\def\ee{\end{equation}}
\def\bes{$$}

\def\ees{$$}
\def\d{{\rm d}}
\def \pa{\partial}

\def\C{{\mathbb C}}

\def\R{{\mathbb R}}

\def\N{{\mathbb N}}

\def\Z{{\mathbb Z}}

\def\res{\mathop{\mathrm{res}}\limits_}

\renewcommand{\theequation}{\arabic{section}-\arabic{equation}}
\makeatletter
\@addtoreset{equation}{section}
\makeatother
\def\tr{\mathrm {Tr}}
\def\le{\left}
\def\ri{\right}

\def\bc{\begin{corollary}}
\def\ec{\end{corollary}}

\def\br{\begin{remark}\rm\small}
\def\1{{\bf 1}}
\def\er{\end{remark}}

\def \scr{\mathscr}

\def\C{{\mathbb C}}
\def\R{{\mathbb R}}
\def\N{{\mathbb N}}
\def\Z{{\mathbb Z}}

\def \eqref#1{(\ref{#1})}

\def\br{\begin{remark}}
\def\er{\end{remark}}
\def\brs{\begin{remarks} \rm\
\begin{enumerate}}
\def\ers{\end{enumerate}\end{remarks}}
\def\bl{\begin{lemma}}
\def\el{\end{lemma}}
\def \bs{\boldsymbol }
\def\bxs{\begin{examps}. \rm\begin{enumerate}}
\def\exs{\end{enumerate}\end{examps}}
\def\bd{\begin{definition}}
\def\ed{\end{definition}}
\def\bp{\begin{proposition}}
\def\ep{\end{proposition}}
\def\be{\begin{equation}}
\def\ee{\end{equation}}
\def\bes{$$}

\def\ees{$$}
\def\Cauchy{\mathbf C}
\def \CC{{\mathcal C}}
\date{}

\date{}

\begin{document}
\fontfamily{cmss}
\fontsize{11pt}{15pt}
\selectfont

\baselineskip 16pt plus 1pt minus 1pt
\begin{flushright}
\end{flushright}
\vspace{0.2cm}
\begin{center}
\begin{Large}
\textbf{\large }\\

\textbf{ 
Algebraic approach to the inverse spectral problem for rational matrices}
\end{Large}
\end{center}
\bigskip
\begin{center}
\begin{large} {M.
Bertola}$^{\ddagger}$\footnote{Marco.Bertola@concordia.ca\\
{\scriptsize Compiled: \today}
}
\end{large}
\\
\bigskip
\begin{small}
$^{\ddagger}$ {\em Department of Mathematics and
Statistics, Concordia University\\ 1455 de Maisonneuve W., Montr\'eal, Qu\'ebec,
Canada H3G 1M8} \\
\end{small}
\end{center}

\bigskip
\begin{center}{\bf Abstract}\\
\end{center}
We  consider the problem of reconstruction of an $n\times n$ matrix with coefficients depending rationally on $x\in \mathbb P^1$ from the data of: (a) its characteristic polynomial and (b) a line bundle of degree $g+n-1$, with $g$ the geometric genus of the spectral curve, represented by a choice of $g+n+1$ points forming a (non-positive) divisor of the given degree. 

We thus provide a reconstruction formula that does not involve transcendental functions; this includes formulas for the spectral projectors and for the change of line bundle, thus integrating the isospectral flows.

 The formula is a single residue formula which depends rationally on the coordinates of the points involved, the coefficients of the spectral curve, and the position of the finite poles of $L$.  We also discuss the canonical bi-differential associated with the Lax matrix and its relationship with other bi-differentials that appear in Topological Recursion and integrable systems.
 \\[10pt]
\setcounter{tocdepth}{5}
\tableofcontents

\section{Introduction and results}
\label{setup}
Let $\CC=\{E(x,y)=0\} $ be a smooth plane algebraic curve, the {\it spectral curve}. This is a polynomial in $x,y$ with $\deg_y E(x,y) =n$, $ \deg_x E(x,y) =m$. 
We denote points on  $\CC$ by $p=(x,y)$, with the two canonical projections indicated by capital letters $X,Y:\CC \to \C$, \ \ \ $X(p)=x$, $Y(p) = y$.   Let $g$ be the genus of $\CC$ (number of points in the Newton polygon).

The goal of the paper is to provide a reconstruction formula to recover  a rational matrix $L(z)$ of size $n\times n$ from the datum of:
\begin{enumerate}
\item The spectral curve  $\CC=\{E(x,y)=0\} $:
\be
\label{spectralcurve}
E(x,y) = \sum_{j=0}^n a_j(x)y^{j}, \ \ \ a_n(x) = \prod_{\ell=1}^K (x-c_\ell)^{\mu_\ell},
\ee
where $a_j(x)$ are otherwise arbitrary polynomials of $x$. Here $E(x,y)=0$ cuts the same locus as the secular equation $\det (y \1 - L(x))=0$. 
\item The choice of a normalization point $z_o\in \C$ and its $n$ pre-images $X^{-1}(\{z_o\}) = \big\{z_o^{(\ell)}, \ \ell = 1,\dots, n\big\} \subset \CC$. It is assumed that $z_o$ is not a branch-point of the $X$ projection, namely, that ${\rm resul}_{w} \big(E(z_o,w), E_y(z_o, w)\big)\neq 0$, where ${\rm result}_w$ denotes the resultant of two polynomials of $w$.   
\item A collection of $g+1$ points $p_o, d_1,\dots d_g$ constituting  a {\it divisor} of degree $g-1$  of the form $\scr D- p_o = \sum_{j=1}^{g} d_j -  p_o $.  
\end{enumerate} 
The line bundle alluded to in the abstract corresponds then to the divisor $\mathcal X = \scr D - p_o + \sum_{\ell=1}^n z_o^{(\ell)}$. 
\paragraph{Some history.}
Reconstruction formulas exist in the literature of integrable systems in terms of Riemann theta functions \cite{ AHH1, AHH2, BabelonBook,Bertola:EffectiveIMRN}. In the rational interpolation approach see \cite{Gohberg1, Gohberg2}.

The goal of this paper is to provide a formula  which is purely algebraic by which we mean that  we do not use any integration (Abel map, Jacobi and inverse Jacobi problem) or  transcendental functions (Riemann Theta functions). In particular it can be {\it implemented on a Computer Algebra System (CAS)}. Depending on the given data, the formula is either algebraic (we may have to solve finitely many  polynomial equations to find the coordinates of points) or rational (if the coordinates of the relevant points are given at the outset).

This is in opposition to the approach using Theta functions which requires several transcendental steps including:
\begin{enumerate}
\item A choice of symplectic basis in homology; this is something that can be done numerically or with the help of CAS's, but becomes rapidly very onerous in term of computational cost;
\item the computation of the matrix of $2g$ periods for each of the $g$ holomorphic differentials; there is a simple choice of holomorphic differentials 
using the Newton polygon, but there is no simple, or natural, choice of canonical symplectic basis in homology. Furthermore, the computation of the periods involves numerically expensive computations since we are to integrate algebraic functions over several sheets of the curve. 
\item The construction and (numerical) evaluation of the theta functions themselves; they are infinite series of the form 
\be
\Theta({\bf z};\bs \tau) = \sum_{{\bf n} \in \Z^g} {\rm e}^{i\pi {\bf n}^t \bs \tau {\bf n} - 2i\pi {\bf n}^t {\bf z}}, \ \ \bs \tau = \bs \tau^t \in {\rm GL}_{g}(\C), \ \ {\bf z} \in \C^g. 
\ee
\item The repeated evaluations of Abel maps of points on the curve; this means the evaluation of the integrals of the normalized holomorphic differentials along paths that must avoid the canonical dissection. 
\end{enumerate}
It should be clear that if an interest is paid to the {\it computability} of the inverse spectral map, this is quickly lost along the way. Moreover diophantine structures are also mangled by this approach.

In the paper of Garnier \cite{Gar2} the problem was tackled. Garnier considers the rational matrix with simple poles
\be
L(x)  = \sum_{\ell = 1}^K \frac {L_j}{x-c_\ell}
\ee
and describes an {\it algorithm} to reconstruct $L(x)$ (up to conjugation by a constant invertible matrix) from the data of the spectral curve (the characteristic polynomial) and $g+n-1$ points. 
This reconstruction scheme was historically important for the development of the Hamiltonian theory of isospectral deformation \cite{AHH1, AHH2} because it turns out that the coordinates of a divisor of degree $g$ in the same equivalence class are the {\em spectral Darboux coordinates} for the Lie-Poisson structure on the space of rational matrices. 

However, the verbal algorithm of Garnier  is not easy to follow and  does not result in an easily computable {\it formula}. We aim at filling this gap with the present short paper.
\paragraph{Main results.}
The formula is explicit, but relies upon an explicit expression for a suitable Cauchy kernel, $\Cauchy_{\scr D - p_o} (p,q)$ (Def. \ref{defCauchy})  which  is the unique differential of the third kind with respect to $p$ with zeros at $\scr D$ and two simple poles at $q,p_o$ of residues $\pm 1$, respectively. Viceversa, with respect to $q$ it is the unique (up to normalization) meromorphic function with poles at $\scr D$ and $q=p$ and a simple zero at $q=p_o$. This is what Weierstrass refers to as "Hauptfunktion" in his lectures \cite{Weierstrass}.

In general such a kernel can be written in several ways that coincide when restricted to the curve. In particular we can write it as 
\be
\Cauchy_{\scr D - p_o} (p,q) = 
 \frac {Q\big((x,y);(z,w)\big)\d x} {E_y(x,y)} , \ \ \ p=(x,y),\ q=(z,w),
\label{Cauchyvague}
\ee
where $Q\big((x,y);(z,w)\big)$ is an appropriate rational function of the variables $x,y,z,w$  (see  Prop. \ref{propCauchy} and \eqref{defOmega}, \eqref{defOmega2}).  With respect to $q=(z,w)$ the dependence is rational and it has poles at $q=p, \scr D$ and simple zero at $q=p_o$.  

The problem of constructing this kernel is discussed in different terms in the last pages of Chapter X of the first part of the Lectures of Weierstrass \cite{Weierstrass}. Some algorithm of construction is implemented in \cite{czni}.  Although algorithms are available, a general formula that applies to any plane curve  is elusive, by which we mean that we can always do the exercise for any curve but we could not find a formula that fits a priori all cases. We can, however, provide a simple expression for \eqref{Cauchyvague} under the assumption that 
\be
\text{The leading and subleading coefficients of $E$ w.r.t $y$ ( or $x$, respectively), do not share any root.}
\label{demonic}
\ee

\begin{figure}
\begin{center}
\begin{tikzpicture}[scale=0.5]
\coordinate(v1) at (1,0);
\coordinate(v4) at (12,8);
\coordinate(v5) at (12,10);
\coordinate(v6) at (11,11);
\coordinate(v7) at (4,11);
\coordinate(v8) at (0,10);

\coordinate (v2) at (8,3);
\coordinate (v3) at (11,5);
\coordinate (p1) at (v2);
\coordinate (p2) at (v3);

\coordinate (ptop) at (11,11);

\draw[white!70!blue, fill] (0,0) to (v1) to (v2) to (v3) to (v4) to (v5) to (v6) to (v7) to (v8) to cycle;
\draw[step=1,black,line width=0.2, dashed] (0,0) grid (13,13);
\draw [black, line width=1,postaction={decorate,decoration={markings,mark=at position 1 with {\arrow[line width=2pt]{>}}}}] (0,0) to node[left]{$\nu$}  (0,13.5);
\draw [black, line width=1,postaction={decorate,decoration={markings,mark=at position 1 with {\arrow[line width=2pt]{>}}}}] (0,0) to node[below]{$\mu$} (13.5,0);
\draw[fill] (p2) node[right] {$(m_{\ell_2},\ell_2)$} circle [radius=0.05];
\draw[fill] (p1) node[right] {$(m_{\ell_1},\ell_1)$} circle [radius=0.05];
\draw[fill]  (ptop) node[above] {$(m_n,n)$}circle [radius=0.05];

\draw (3.5,0) to node[above, pos = 0.2]{$\scr L$} (13, 6.3333);

\end{tikzpicture}
\end{center}

\caption{An example of Newton polygon; the assumption \eqref{demonic} (predicated for the variable $y$ in this example) means that the polygon, for any translation in $x$, has a vertex at $(0,n)$ or $(0,n-1)$ (as depicted). }
\label{NewtonPolygon}
\end{figure}
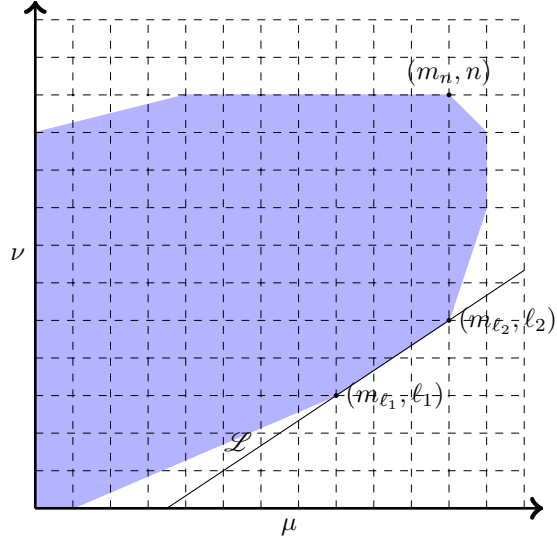
In particular this includes the case where $E(x,y)$ is monic in either $x$ or $y$. 
The formula  is  in Prop. \ref{propCauchy}, where $\Omega$ ibidem is given by \eqref{defOmega} or \eqref{defOmega2}, depending on which case we are considering.
Geometrically these assumptions have the following interpretation: if $\CC$ denotes the desingularization of the curve at the infinite points, then, for example,  the assumption that $E$ is monic with respect to $y$ implies that all poles of the $Y$-projection are also poles of the $X$-projection; if $E = x^m + \mathcal O(x^{m-1})$, similarly, all poles of $X$ are also poles of $Y$. One could, however, have additional poles for the projections: for example in the second case, $Y$ may have poles of any order at points where $X$ is regular. These correspond to the poles of the matrix $L(x)$. 

Under the general assumption  \eqref{demonic} instead, the geometric meaning is as follows. If $E = a_n(x) y^n + a_{n-1}(x) y^{n-1} + \dots$ and $a_{n}, a_{n-1}$ have non-zero resultant, then above the zeros $x = b$  of $a_n(x)$ there is a single pole of $Y$ of order exactly equal to the multiplicity of the root $b$ in $a_n$.  In particular these cannot be branch-points of the $X$ projection in the de-singularization.
See Lemma \ref{lemmaOmega} and Prop. \ref{propCauchy}.

We can then encapsulate the formula in the following theorem.

\bt
\label{main}
Let $\scr D = \sum_{\ell=1}^g (d_{\ell x}, d_{\ell y})$ be a non-special divisor  of degree $g$ on $\CC$. 
Let $p_o  = (p_{ox}, p_{oy})\in \CC$ and $z_o\in \C$ a complex value, not branchpoint of the $X$ projection of the spectral curve \eqref{spectralcurve}. Let $z_o^{(\ell)} = (z_o, w_o^{(\ell)}) $ denote the pre-images of $z_o$ on $\CC$. 
Finally, define the matrix $L$ with entries 
\bea
L_{ab}(x):=& (x-z_o)^2 \hspace{-15pt}\sum_{c\in \{d_{\ell y},\ell=1..g,\atop   \ w^{(a)}, w^{(b)} , \infty\}}\hspace{-10pt}\res{\blue{y}=c} 
 \frac {
 Q\big( z_o^{(a)};(x, \blue{y}) \big) \,\blue{y}\,
Q\big((x, \blue{y}); z_o^{(b)} \big) 
 d \blue{y} }{E_y(z_o, w_o^{(a)})E(x,\blue{ y}) }.
\eea
 Then the characteristic polyomial $\det (y\1 - L(x))=0$ cuts the same locus as $E(x,y)=0$. The divisor class $\scr X = \scr D - p_o + \sum_{j=1}^n z_o^{(j)}$ determines a line bundle of degree $g+n-1$ to which the components of the left eigenvector belong. 
 \et
See Theorem \ref{reconstruction_theorem}.
We provide numerical examples in Section \ref{examples}.
%
\paragraph{Spectral projectors and spectral bidifferential.}

Another interesting formula is the following one: let $(x,y)\in \CC$ and consider the  spectral projector on the eigenspace corresponding to the eigenvalue $y$ of $L(x)$: this was considered previously in \cite{Bertola:EffectiveIMRN, Dubrovin}. This can be written explicitly as follows (a simple exercise of linear algebra):
\bea
\Pi(x,y):= \frac {\wt {\le(y\1 - L(x)\ri)}}{\tr {\wt {\le(y\1 - L(x)\ri)}}} = \frac {\wt {\le(y\1 - L(x)\ri)}}{P_y(x,y)}, \ \ \ \tr \le(\Pi(x,y)\ri) \equiv 1, \nn\\
P(x,y):= \det \le(y \1 - L(x)\ri). \label{projector}
\eea
The tilde above the matrix denotes the {\it classical adjugate matrix}, namely, the matrix of co-factors, transposed. In \cite{Dubrovin}, formula (1.10) an equivalent expression is provided for the adjugate matrix using the Cayley-Hamilton theorem. 
In \cite {Bertola:EffectiveIMRN} the following canonical bi-differential associated to the Lax matrix was introduced:
\be
\label{defbidiff}
B(p,q):= 
\frac {\tr\Big( \Pi(x,y) \Pi(z,w)\Big)}{(x-z)^2} \d z \d x, \ \ p = (x,y), \ q = (z,w).
\ee
This is the same considered in \cite{Dubrovin} (1.11). 
The above expression lifts to a well--defined  bi-differential on $\CC$ with the property that 
\begin{enumerate}
\item it is symmetric under the swap $p\leftrightarrow q$.
\item With respect to $p$ it is a second-kind differential with a single double order pole at $p=q$. 
\end{enumerate}
We refer to \eqref{defbidiff} as the {\it spectral bidifferential}. 
These types of bi-differentials are essential in the development of the theory of Riemann surfaces and topological recursion \cite{EO}.  The classical object of similar nature is the {\it fundamental bi-differential} (sometimes called {\it Bergman} bi-differential), which is the unique bi-differential normalized in such a way that all the $a$--periods vanish, for a choice of maximally Lagrangian subspace in the homology of the curve; see \cite{Faybook}, Ch. II. 

In contrast,  the spectral bidifferential $B(p,q)$ \eqref{defbidiff} does not have a discernible normalization. It can be written however  {\it in two ways}  by relating it to the Cauchy kernel \eqref{Cauchyvague} or to the Szeg\"o kernel (see Section \ref{Szegosection}). 

\bc
\label{bidiffcoroll}
The bidifferential \eqref{defbidiff} is given by 
\be
B(p,q) =  \Cauchy_{\scr D-p_o} (p;q)\Cauchy_{\scr D-p_o} (q;p).
\ee
\ec
In different form, Corollary \ref{bidiffcoroll} appeared in \cite{Bertola:EffectiveIMRN}.
The proof is in Section \ref{proofcorollary}. 
We comment in Section \ref{Szegosection} on the alternative expression of the bi-differential in terms of the Szeg\"o kernel, although it is outside of the spirit of the present paper being that the Szeg\"o kernel is a transcendental object. 

\paragraph{Correlation functions in integrable systems.}
In \cite{Dubrovin}, building up on the main idea of \cite{BDY}, see \cite{BE} as well, the following formula was derived (we re-cast in a slightly different, but equivalent form), see eq. (1.12) in loc. cit..

Consider the {\it multi-point correlators} defined by  ($N\geq 3$)
\bea
W_N(p_1, \dots, p_N):= 
\frac{(-1)^N}{N} \sum_{\sigma\in 
S_N}  
\frac {\tr\,\Big[\Pi(p_1)\cdots \Pi(p_N)\Big]}{(x_{\s_1} - x_{\s_2}) (x_{\s_2}- x_{\s_3}) \cdots (x_{\s_N} - x_{\s_1})}\ ,    \ \ \ p_j= (x_j, y_j).
\eea

We have 
\bc
\label{polydiffcoroll}
For $N\geq 3$ 
\be
W_N(p_1, \dots, p_N) \prod_{j=1}^{N} \d x_j = \frac{(-1)^N}{N} \sum_{\sigma\in 
S_N}  \Cauchy_{\scr D-p_o}(p_{\s_1}, p_{\s_2}) \cdots \Cauchy_{\scr D-p_o}(p_{\s_N}, p_{\s_1})
\ee
For $N=2$  see \eqref{defbidiff} and Corollary \ref{bidiffcoroll}. 
\ec
The proof is in Section \ref{proofcorollary}.

The relevance of these formulas stems from the fact that $W_N$'s are the so-called multi-correlators of a multi-KP tau function on one side (see the Main Lemma therein), and on the other side they are expressed in terms of the derivatives of the Riemann Theta function evaluated at a particular point of the Jacobian, see Corollary 1.3 in loc. cit. In our notation this reads
\be
\sum_{\ell_1,\dots, \ell_N=1}^{g} \omega_{\ell_1}(p_1)\cdots \omega_{\ell_N}(p_N) \frac {\pa^N}{\pa z_1\cdots \pa z_N} \ln \Theta_{\mathcal X}(0) = 
W_N(p_1, \dots, p_N) \prod_{j=1}^{N} \d x_j 
\ee

\section{Spectral  data and reconstruction}
According to the classical results \cite{Gar2, AHH1, AHH2}
a rational matrix $L(x)$ can be recovered, up to conjugation by a constant matrix, from the data of:
\begin{enumerate}
\item The {\it spectral curve} $\CC$, i.e., its characteristic polynomial, understood as cutting a locus in the $(x,y)\in \C^2$ surface. 
\item A divisor $\scr X$, up to linear equivalence, of degree $g+n-1$, where $g$ is the genus of the de-singularization of $\CC$ and $n$ is the dimension of $L(x)$. Equivalently we can think of this divisor as a line bundle $\scr X$ of degree $g+n-1$. Equivalently again, the holomorphic sections of such a line bundle are identified with the vector space of meromorphic functions whose divisor exceeds  $\scr X$. We will use the symbol $\scr X$ to indicate both the divisor and the line bundle. 
\end{enumerate}
The idea that a non-abelian object (a matrix $L(x)$) can be recovered from abelian data (spectral curve and line bundle) is the core of the more general  process of  abelianization of Hitchin \cite{Hitchin} in the context of Higgs bundles. 

Since the divisor $\scr X$ is only determined up to linear equivalence, the expedient way of fixing it proceeds as follows; 
\begin{itemize}
\item choose a {\it normalization point} $z_o\in \C$ as explained in the introduction (not on the branching locus) and denote by $z_o^{(a)} = (z_o, w_o^{(a)})$ , $a=1,\dots, n$ its $n$ preimages on $\CC$ under the $X$ projection, namely the $n$ solutions of  $E(z_o, w)=0$.  We choose it not a branch-point for simplicity.
\item We can choose $n-1$ points arbitrarily. Thus we can define $n$ {\it effective}\footnote{Namely, consisting of sum of points only with positive multiplicities.} divisors of degree $g$ by the relation 
\be
\label{divisors}
\scr X  \equiv \scr D_g^{(\ell)} + \sum_{j\neq \ell} z_o^{(j)}.
\ee 
\item Finally, we can represent $\scr X$ as a unique {\it non-effective} divisor of the form 
\be
\scr X \equiv \scr D - p_o + \sum_{j=1}^{n} z_o^{(j)}. 
\ee
\end{itemize}
From the definition we obtain that these divisors all satisfy the relation
\be
\scr D_g^{(\ell)} - z^{(\ell)}_o \equiv \scr D_g^{(m)} - z^{(m)}_o ,\ \ \ \ell, m=1,\dots, n,
\label{reldivisors}
\ee
where the equivalence is the linear one. Namely there are meromorphic function $F_{\ell,m}(p)$ such that\footnote{For a function $F$ or any meromorphic section of a line bundle, we denote by $\div(F)$ its divisor, namely, the formal sum of points counted with the order of $F$ at those points (multiplicity of zero or minus order of the pole). }
\be
\div F_{\ell,m} = \scr D_g^{(\ell)} - z^{(\ell)}_o -\scr D_g^{(m)} + z^{(m)}_o.
\ee
We will write these functions explicitly in Remark \ref{cobonudary}.

\paragraph{The Cauchy kernel.}
Let $\scr D$ be a non-special divisor of degree $g$,  $p_o$ another point not belonging to $\scr D$.
We recall that ``non-special'' means that the $\ell (\scr D) = 1$ (where $\ell(\scr D)$ is the dimension of the vector space of rational functions, $F$,  with pole divisor exceeding $-\scr D$, $\div(F)\geq - \scr D$), namely  (if $\scr D$ is positive) the only such function is the constant.
 \bd 
\label{defCauchy}
The Cauchy kernel is the unique kernel (i.e. depending on two variables $p,q\in \CC$) which is a meromorphic differential with respect to the first variable $p$,  a meromorphic function with respect to the second variable $q$ and with the properties:
\bea
\div_p \Cauchy_{\scr D-p_o} (p,q) \geq \scr D-(p_o)-( q);\ \ 
\\
\div_q \Cauchy_{\scr D-p_o} (p,q) \geq -\scr D+(p_o)- (p);\ \ \\
\res{p=q}  \Cauchy_{\scr D-p_o} (p,q)=1 = -\res{p=p_o}  \Cauchy_{\scr D-p_o} (p,q)
\eea
\ed
The existence and uniqueness follows from the Riemann-Roch theorem as a consequence of the non-specialty of the divisor $\scr D$.  
The Cauchy kernel described above is a classical object that appears in \cite{BehnkeStein,Faybook,Rodin,Zvero}.
The key fact upon which the whole construction relies is the following
\bp
\label{Cauchyrational}
The Cauchy kernel $ \Cauchy_{\scr D-p_o} (p,q)$ depends  {\bf rationally} on the coordinates of the points in $\scr D-p_o$ and the coefficients of the plane curve $E(x,y)$.  
\ep
The proof relies on the explicit formula which is given in Section \ref{formula} below. 
\br
\rm 
\label{cobonudary}
The function that realize the linear equivalence  between the spaces of global section of the sheaf $\mathcal O( \scr D_\ell - z_o^{(\ell)})$ \footnote{By the notation $\mathcal O(\scr V)$ for a divisor $\scr V$ we mean the sheaf of functions $F$ such that $\div(F)+\scr V\geq 0$.} for different $\ell$'s is given by 
\be
F_{\ell,m}(p) =\frac{ \Cauchy_{\scr D-p_o} ( z_o^{(m)}, p )}{ \Cauchy_{\scr D-p_o} ( z_o^{(\ell)}, p )}.
\ee
Multiplication by $F_{\ell, m}$ maps $H^0(\mathcal O( \scr D_m - z_o^{(m)}))$  to $H^0(\mathcal O( \scr D_\ell  - z_o^{(\ell )}))$
Similarly, the functions
\be
\label{fell}
f_{\ell} (p):={ \Cauchy_{\scr D-p_o} (z_o^{(\ell)}, p)}, \ \ \ell = 1, \dots , n
\ee
are the meromorphic functions  spanning $H^0\le(\mathcal O\big ( \scr D - p_o + \sum_{j=1}^{n} z_o^{(\ell)}\big)\ri)$.
Moreover the expression \eqref{fell} shows that all the following divisors of degree $g-1$ are linearly equivalent (here $\sim$ denotes linear equivalence of divisors\footnote{We recall that two divisors $\scr V, {\scr V'}$ are {\it linearly equivalent} if there exists a rational function $F$ such that $\div (F) = \scr V - \scr V'$.})
\be
\label{equivalence}
\scr D-p_o \sim \scr D_{\ell} - z_o^{(\ell)}, \ \ \ \ell = 1,\dots, n. 
\ee
The Serre-dual bundle $\mathcal K_\CC(-\scr D+p_0)$ is spanned by 
\be
\label{phiell}
\varphi_\ell = \Cauchy_{\scr D-p_o}  (p,z_o^{(\ell)}), \ \ \ell = 1,\dots , n. 
\ee
\hfill $\blacktriangle$
\er

\paragraph{The Eigenvector matrix.}
Let us define the $n\times n$ matrix $\Psi(z_o;x)$ with entries.
\be
\label{PsiCauchy}
\Psi_{ab}( z_o; x):= \Cauchy_{\scr D - p_o }(z_o^{(a)}, x^{(b)}).
\ee
Note that the entries $\Psi(z_0;x)$ are differentials with respect to $z_0$ and algebraic functions with respect to $x$. 
The following Lemma appears repeatedly in the literature; possibly the first instance is in \cite{Korotkin05}, where, however, instead of the Cauchy kernel the author uses the Szeg\"o kernel, which is expressed in terms of Theta functions (hence not algebraic).  See Remark \ref{remdima}.
\bl
\label{lemma31}
The inverse to the matrix $\Psi(z_o;x)$ is the matrix $\Psi(x;z_o) \frac{(x-z_o)^2} {\d x \,\d z_o} $. 
\el
{\bf Proof.}
The proof will follow once  we prove the  claim that
\be 
\sum_{c=1}^n  \Cauchy_{\scr D - p_o }(z_o^{(a)}, x^{(c)})
 \Cauchy_{\scr D - p_o }(x^{(c)}, z_o^{(b)}) = \delta_{ab} \frac {\d x \d z_o}{(x-z_o)^2}.
\ee
To prove the assertion we inspect the left hand side; since it is invariant under permutation of the pre-images of $x$, it must descend to  a meromorphic differential on $\P^1$. There are no poles at the image $X(\scr D)$ because the poles in the Cauchy kernel cancel againts the zero of the other Cauchy kernel in the expression (in the variable $x$). Ditto for the pole at $x^{(c)} = p_o$. Thus the only possible remaining pole is for $x^{(c)} = z_o^{(a)}, z_o^{(b)}$, both projecting to $x = z_o$. For $a\neq b$ the pole at $x=z_o$ must be simple. But since there are no non-trivial differentials on $\P^1$ with only one simple pole, we conclude that the expression vanishes for $a\neq b$. For $a=b$ there is a double pole at $z=z_o$ with unit leading coefficient if we evaluate the expression in the projection coordinate $X(z_o^{(a)}) = z_o, \ \ \ X(x^{(c)})=x$. The proof follows.
\QED
The $\ell$-th row  of $\Psi$ consists of  the evaluations on the different sheet of the unique (up to multiplicative scalar) meromorphic function $f_\ell$  with divisor properties 
\be
\div(f_\ell)\geq -\scr D- z_o^{(\ell)}  + p_o,
\ee
appearing in \eqref{fell}. 
The span of the functions $f_1,\dots f_n$ coincides with the linear space $H^0(\mathcal O(\scr X))$ i.e. the line bundle of degree $g+n-1$ given as initial datum.
Viceversa, the $\ell$-th column of $\Psi^{-1} $ is the evaluation (divided by the pullback of $\d z$) of the unique meromorphic differential $\varphi_\ell$ with divisor properties 
\be
\div(\varphi_b)\geq \scr D - z_o^{(\ell)} - p_o 
\ee
and given by \eqref{phiell}. 
These are sections of $\scr X$ and the \replace{s}{S}erre dual $\scr X^\vee \otimes \mathcal K$. 
With this in mind we conclude with the, by now obvious, theorem below. 
\bt
\label{LaxCauchy}
The Lax matrix $L(x)$ given by 
\be
L(x) :=\frac{(x-z_o)^2}{\d x\d z_o}  \Psi(z_o;x) \mathbb Y(x) \Psi(x;z_o), 
 \ \ \mathbb Y(x):= {\rm diag} (y_1(x), \dots y_n(x))
\ee
or, equivalently, 
\be
L_{ab}(x) = \frac{(x-z_o)^2}{\d x \d z_o} \sum_{c=1}^n  
 \Cauchy_{\scr D - p_o }(z_o^{(a)}, x^{(c)})  y^{(c)} 
  \Cauchy_{\scr D - p_o }(x^{(c)}, z_o^{(b)}), 
\ee
where $y^{(c)} $ are the $n$ solutions of the polynomial equation $E(x,y)=0$ and $x^{(c)} = (x, y^{(c)} )$. This 
is a rational matrix of $x$ whose characteristic polynomial $\det \le(y \1 - L(x)\ri)=0$ cuts the same affine locus as $E(x,y)=0$. 
\et
The above theorem does not provide yet a computable formula because it requires the solution of the algebraic equation $E(x,y)=0$. However, in the next Section \ref{formula}, it will appear how the formula is implementable and leads to the formul\ae\ in Theorem \ref{main}.

\br
\label{remdima}
In \cite{Korotkin05} Korotkin constructs the solution of the inverse problem of  {\it quasi-permutation monodromy} Riemann-Hilbert problem.
The relationship between that problem and the spectral problem is as follows. The data in loc. cit. are a collection of branchpoints $\{b_{\ell} \} \subset \P^1$ and a representation of the fundamental group $J:\pi_1(\P^1\setminus \{b_\ell\}, z_o)\to {\rm SL}_n(\C)$. The representation is such that all matrices $J_\gamma$ have the structure of a permutation matrix, $\Pi_\gamma$, multiplied by an arbitrary invertible diagonal matrix, i.e. a {\it quasi-permutation}. 
 The solution in loc. cit. is constructed in terms of the Szeg\"o kernel of the ramified cover, $\CC$, of $\P^1$ with a meromorphic projection $X:\CC \to \P^1$ such that its  branching structure is equal to the permutation structure of the monodromy; namely, the author constructs a matrix $\Psi(x)$  such that\footnote{The notation $x^\gamma$ in the argument of an analytic function means the analytic continuation along the closed path $\gamma$.}
 \be
\Psi(x^\gamma) = \Psi(x) J_\gamma, \ \ \gamma \in \pi_1(\P^1\setminus \{b_\ell\}, z_o), \ \ \ \Psi(z_o)=\1, \ \ \ \det \Psi(x)\equiv 1
 \ee
 This matrix can then be used as the eigenvector matrix as follows; choose an arbitrary meromorphic function $Y: \CC \to \P^1$, so that the two projection satisfy a polynomial relation $E(X(p), Y(p))\equiv 0$. Then the diagonal matrix $\mathbb Y(x)$ whose entries are the evaluations of $Y$ of the  pre-images $X^{-1}(\{x\})$ satisfies an adjoint version of the same Riemann Hilbert problem:
 \be
 \mathbb Y(x^\gamma) = \Pi_\gamma \mathbb Y(x)\Pi^{-1}_\gamma.
 \ee
 It is then seen that $L(x):= \Psi(x) \mathbb Y(x) \Psi(x)^{-1}$ is a single--valued meromorphic function of $x\in \P^1$,  and this  is our Lax matrix. 
 
 This paper borrows heavily on the main idea of the construction, but replaces the Szeg\"o kernel, which is a {\it transcendental} object constructed in terms of Theta functions, with the Cauchy kernel subordinated to the choice of divisor, which is instead an {\it algebraic} object. 
\er

\section{The Cauchy kernel and reconstruction formula}
\label{formula}
It is possible to write expressions for $ \Cauchy_{\scr D-p_o} (p,q)  $ in terms of $\Theta$-functions \cite{Bertola:EffectiveIMRN}   but here we want to express it as explicitly as possible in terms of a rational expression involving the plane model of the curve. 

The notation is then as follows: the Newton polygon is the set
\be
\mathcal N:= {\rm Hull} \le\{ (i,j)\in \N^2: \ \ \ {\rm Coeff}_{x^i y^j}(E) \neq 0\ri\}
\ee
where ${\rm Hull}$ denotes the (closure of the)  convex hull and ${\rm Coeff}$ the coefficient of a particular monomial in a  polynomial. The genus coincides with the number of integer points in the interior $\dot {\mathcal N}$. With the notation $\alpha = (\alpha_1,\alpha_2)\in \N\times \N$, we can write a  basis of holomorphic differentials as follows:
\be
\eta_\alpha(p):= \frac{ x^{\alpha_1} y^{\alpha_2} \d x}{E_y(x,y)}, \ \ p = (x,y), \ \ (\alpha_1, \alpha_2)\in \dot {\mathcal N}_<:= \dot {\mathcal N} - (1,1),
\ee
where $\dot{\mathcal N}$ denotes the integer lattice points in  the interior of the Newton polygon and the subscript $_<$ is a short-hand to denote the shift by $-(1,1)$. We will call this the {\it diminished polygon}. 
We work with a generic simple divisor $\scr D$ and denote its points with $d_\ell = (d_{\ell x}, d_{\ell y}), \ \ \ell=1,\dots, g$. With reference to the expression \eqref{spectralcurve} for $E(x,t)$
In order to arrive at the most explicit formula, we now enforce the assumption  \eqref{demonic}  with respect to the variable $y$, for definiteness
\bea
\Omega(p;q,p_o):= &\frac{ E(z, y )}{(x -z)(y-w)}  - \frac{ E(p_{o x}, y)}{(x - p_{ox})(y-p_{oy})}
 + y^{n-1} \le(\frac {a_n(z) - a_n(x)}{z-x}-  \frac {a_n(p_{ox}) - a_n(x)}{p_{ox}-x}  \ri)\nn
  \\
  &p=(x,y), \ \ q = (z,w), \ \ p_o = (p_{ox}, p_{oy}).
  \label{defOmega}
\eea
Note that in the case the curve is monic with respect to $y$ ($a_n\equiv 1$)   the last term is absent.  The case when the assumption \eqref{demonic} is with respect the $x$-variable is similar, with the roles of $x\leftrightarrow y$ swapped in the above formula. 
\br
Just for  reference, in the case 
\be
E(x,y) = b_m(y) x^m + b_{m-1}(y) x^{m-1} + \sum_{\ell=0}^{m-2} b_\ell(y) x^\ell, \ \ {\rm result}_y (b_m, b_{m-1}) \neq 0, 
\ee
the formula \eqref{defOmega} becomes
\bea
\label{defOmega2}
\Omega(p;q,p_o):= &\frac{ E(x, w )}{(x -z)(y-w)}  - \frac{ E(x,p_{o y})}{(x - p_{ox})(y-p_{oy})}
 + x^{m-1} \le(\frac {b_m(w) - b_m(y)}{w-y}-  \frac {a_n(p_{ox}) - b_m(y)}{p_{oy}-y}  \ri)\nn
  \\
  &p=(x,y), \ \ q = (z,w), \ \ p_o  = (p_{ox}, p_{oy}). 
\eea
\er
\bl
\label{lemmaOmega}
The expression 
\be
\Omega(p;q, p_o) \frac {\d x} {E_y(x,y)}=- \Omega(p;q, p_o) \frac {\d y} {E_x(x,y)}
\label{thirdkind}
\ee
 restricted to the spectral locus $E(x,y)=0 = E(z,w)= E(p_{ox}, p_{oy})$  is a third-kind differential of $p$ with only two poles on $\CC$ at $p=q, p_o$ and residues $\pm 1$, respectively. 
\el 
This is an exercise which we address in Appendix \ref{exercise}.
We point out, for  reference,  that 
the function $ \Omega(p;q, p_o)$ in \eqref{defOmega} has the properties 
\begin{enumerate}
\item with respect to $x$  it is a rational function with only two simple poles at $x =z, p_{ox}$
\item with respect to $y$ it is a polynomial of degree $n-1$.
\item with respect to $z$ it is a polynomial of degree $\deg_x E -1 $. 
\item with respect to $w$ it  is a rational function with only simple poles at $w = y, p_{oy}$. 
\end{enumerate}
For \eqref{defOmega2} the roles of the $X$ and $Y$ projections, in all variables, are swapped.

The above expression emphasizes the dependence on the coordinates of $p,q$; note that $q, p_o$ play a completely symmetrical role (up to overall sign), but in the construction one of them is thought of as a parameter, whence the shifted emphasis. 
\bp
\label{propCauchy}
The Cauchy kernel subordinated to $\scr D - p_o$ at the points $p= (x,y),\  q = (z,w)$  is given by the expression 
\bea
\label{Cauchyexp}
\Cauchy_{\scr D-p_o}(p,q) =& 
\frac {Q\big((x,y); (z,w) \big)\d x}{E_y(x,y)},\nn\\
Q\big((x,y); (z,w) \big):=&  \Omega(p;q,p_o) - \sum_{\alpha \in \dot {\mathcal N}_{<}} Q_\alpha x^{\alpha_1} y^{\alpha_2} 
\eea
where $\Omega$ is the rational function of $p,q,p_o$ in \eqref{defOmega}. 
The coefficients $Q_\alpha$ are found by imposing that the bracketed expression vanishes, as a function of $p$, for $p\in \scr D$, namely, by solving the linear system 
\be
 \sum_{(\alpha_1,\alpha_2) \in \dot {\mathcal N}_{<}} 
 Q_\alpha((z,w);(p_{ox}, p_{oy}) )  d_{\ell x} ^{\alpha _1} d_{\ell y}^{\alpha_2}=\Omega(d_\ell;q, p_o),  \ \ \ \ell = 1, \dots, g, \ \ \ q = (z,w), 
 \label{Qsystem}
\ee
with $\Omega $ as in \eqref{defOmega}.
\ep
\br
From the system \eqref{Qsystem} it is manifest that the functions $Q_\alpha( q; p_o)$ are odd $Q_{\alpha}(q;p_o) = - Q_{\alpha}(p_o;q)$ and moreover they have poles w.r.t $q$ ($p_o$) when $q\in \scr D$ and a zero when $q= p_o$. 
\er
{\bf Proof.}
The proposition is  self-evident, since the term containing $\Omega$ is already a third-kind differential, and the extra term consists of holomorphic differentials which are chosen to fix the divisor of zeros (with respect to $p$) of the differential.  We thus have only to discuss the solvability of the system \eqref{Qsystem}. It is at the core of the proof of the Riemann--Roch theorem that the co-rank of the $g\times g$ matrix of the system $\Delta$ indexed by $\ell\in [1,\dots, g]$ and $\alpha \in \dot{\mathcal N}_<$
$$
\Delta_{\ell, \alpha}  = [d_{\ell x}^{\alpha_1} d_{\ell y}^{\alpha_2}]_{\ell=1\dots g \atop \alpha \in \dot{\mathcal N}_<}
$$
is, by definition, the index of specialty of $\scr D$. Thus the non-specialty of $\scr D$ is equivalent to the invertibility of $\Delta$ (the {\it Brill-Noether} matrix).
\QED

 Note also that it is irrelevant, for the reconstruction formula of the next section,  whether $\Omega$ is given by \eqref{defOmega} or \eqref{defOmega2}.


We should view the $g\times g$ matrix $\Delta = [d_{\ell}^\alpha]_{\ell=1\dots g \atop \alpha \in \dot{\mathcal N}_<}$ of this system as a sort of generalized Vandermonde matrix. The system also manifests the fact that the coefficients $Q_\alpha(z,w) = Q_{\alpha} (q; p_o)$ are rational in the coordinates of $q, p_o$ and have poles for $q\in \scr D$ and zero for $q=p_o$. 

We  finally   observe that the expression is a rational expression in the coordinate of all points involved.

\subsection{Reconstruction formula}

\bt
\label{reconstruction_theorem}
Let $z_o^{(a)}  = (z_o, w_o^{(a)}), \ \ a=1,\dots, n$ be points on the spectral curve projecting to the same point $X( z_o^{(a)})=z_o$ of the complex plane. 
Then the Lax matrix with spectral locus $E(x,y)=0$ and divisor class $\scr D-p_o + \sum_{k=1}^n z_o^{(k)}$ is given by the residue formula below
\bea
\label{preLaxsum}
L_{ab}(x):=&
 (x-z_o)^2\hspace{-10pt}
 \sum_{c\in \{d_{1 y},\dots, d_{g y},  p_{oy}\atop   \ w_o^{(a)}, w_o^{(b)}, \infty\}}\hspace{-10pt} \res{\blue{y}=c} 
 \frac {
 Q\big((z_o,w_o^{(a)} );(x,\blue{y})  \big) 
 \blue{y} \,
 Q\big((x,\blue{y}); (z_o,w_o^{(b)} ) \big) 
\d \blue{y}}{E(x,\blue{ y}) E_y(z_o,w_o^{(a)} )}. 
\eea
where $Q\big( (x,y); (z,w)\big)$ is given in \eqref{Cauchyexp}.
The  normalization is  such  that $L(z_0) $ is diagonal. 
\et
{\bf Proof.}
The formula is the implementation of Theorem \ref{LaxCauchy}, together with the use of Cauchy's residue theorem, as we now see.
Let $x^{(k)}  = (x,y^{(k)})$ with  $y^{(k)}$  the (generically distinct) $n$ solutions of $E(x,y)=0$. Consider the sum over all sheets
\be
\sum_{k=1}^n  
 \frac { Q\big((z_o^{(a)};(x,y^{(k)}) \big)\,
  y^{(k)} \, Q\big((x,y^{(k)});z_o^{(b)} \big) } {E_y(x,y^{(k)}) E_y(z_o, w_o^{(a)} )}.
\ee
Each term is just the residue at $y =y^{(k)}$ of the expression
\be 
 \frac { Q\big((z_o^{(a)};(x,y) \big)\,
  y \, Q\big((x,y);z_o^{(b)} \big) \d y } {E(x,y) E_y(z_o, w_o^{(a)} )}.
 \ee
As a differential  of $y$ this is a meromorphic differential on $\mathbb P^1$  with the only other poles at $y = d_{\ell y}, \ell = 1,\dots g$ and $y = p_{oy}, w_o^{(a)}, w_{o}^{(b)},  \infty$. Whence the formula in the theorem follows immediately from the fact that the sum of all residues of a  rational function  on $\mathbb P^1$ is zero. \QED

 We point out that the evaluation of the residues is at points that {\bf do not depend} on $x$ and are fixed as part of the data. Moreover 
 \bes
 \text{ the expression \eqref{preLaxsum} is a rational function of $x$ and it is rationally computable.}
 \ees
 By "rationally computable" we mean  that there is no need to solve an algebraic equation as long as the coordinates of the points in the constructions are given. Thus, for example, if $E(x,y)$ is with integer coefficients and all the points $z_o^{(\ell)}, \scr D, p_o$ are with rational coordinates, then the matrix $L(z)$ will also be over $\mathbb Q$. 
We also show 
\bl
\label{nopole}
The expression for $L_{ab}(x)$ \eqref{preLaxsum} has no poles in $x$ at the $x$--projections of the points in $\scr D$, namely, $x = d_{\ell x}$,  $\ell = 1,\dots, g.$
\el 
{\bf Proof.}
The expression \eqref{preLaxsum} has, {\it prima facie}, poles for $y= d_{\ell,y}$, $\ell =1,\dots, g$. Looking at \eqref{Qsystem} these would produce poles for $x= d_{\ell,x}$.  However, as $(x,y)\to (d_{\ell,x}, d_{\ell,y})$ the pole of the second bracket is cancelled by the zero of the first bracket. Hence we are forced to conclude that resulting expression  is analytic at $x=d_{\ell,x}$.  
\QED

If $x=\infty$ is not a branchpoint (this happens if none of the sides of the Newton polygon that face north-east  has slope $-\frac p q$ with $p>1$) then we can move the normalization point $z_0$ to infinity.

The poles of $L(x)$ will be only at the poles of $Y:\CC \to \mathbb P^1$, and hence at the zeros of the leading coefficient $a_n(x)$ of $E(x,y) = y^n a_n(x) + \dots..$.  Clearly, thus, if $E(x,y) = y^n + \dots$,  $L(x)$ will be a polynomial matrix. 
\subsection{Change of divisor}
The divisor $\scr D-p_o$ is called ``dynamical divisor'' in that, under the isospectral flows, it depends on time(s) linearly, when expressed as a point in the Jacobian variety $\mathbb J(\CC)$. We can thus integrate all these flows at once by expressing the map corresponding to the change of divisor from $\scr D$ to $\wt {\scr D}$ (both of degree $g$)\footnote{It is unnecessary, in view of Jacobi inversion theorem, to also change $p_o$.}

Suppose thus that we have the Cauchy kernels  in the form \eqref{Cauchyvague} for two different divisors, 
\bea
\label{cauchyvague1}
\Cauchy_{\scr D-p_o}(p;q) = \frac {Q\big((x,y);(z,w)\big) \d x} {E_y(x,y)},\\ 
\label{cauchyvague2}
\Cauchy_{\wt {\scr D}-p_o}(p;q) = \frac {\wt Q\big((x,y);(z,w)\big) \d x} {E_y(x,y)},\\ 
\eea
Then the Lax matrices $L(x), \wt L(x)$ with the same spectral curve but with the different divisor data will be related by 
\be
\wt L(x)  = P(x)  L(x) P^{-1}(x), \
\ee
where the transition matrix $P(x) = P_{\scr D, \wt {\scr D}}(x)$ effecting the shift of the spectral divisors is given by 
\be
[P_{\scr D, \wt {\scr D}}(x)]_{ab} = \frac{(x-z_o)^2}{\d x \d z_o} \sum_{\ell=1}^n \Cauchy_{\wt{\scr D}-p_o}(z_o^{(a)}, p^{(\ell)})
\Cauchy_{{\scr D}-p_o}( p^{(\ell)}, z_o^{(b)}),\ \ \ \ 
P_{\scr D, \wt {\scr D}}(x)^{-1} = P_{\wt{\scr D},  {\scr D}}(x)
\ee
By the same argument used in Lemma \ref{lemma31}, the above expression evaluates to a rational function of $x$ with poles at $X(\scr {\wt D})$ only.  Using the two  expressions \eqref{cauchyvague1}, \eqref{cauchyvague2} and using the same observation as in the proof of Theorem \ref{reconstruction_theorem} we convert to a finite sum of residues in the $y$--variable:
\bea
[P_{\scr D, \wt {\scr D}}(x)]_{ab} =&(x-z_o)^2 \sum_{c \in \{ 
 \wt d_{y j}, j=1\dots g\atop
z_o^{(a)}, z_o^{(b)}, \infty \}}
\res{\blue{y}=c} 
 \frac {
\Big[\wt Q((z_o^{(a)};(x,\blue {y})) \Big] 
\Big[Q((x,\blue{ y}) ;z_o^{(b)})  \Big]
\d \blue{y} }{E(x,\blue{ y}) E_y(z_o, w_o^{(a)})}.
\eea

\subsection{The spectral projector and bi-differential}
\label{proofcorollary}
By definition, the spectral projector  is the outer product of the right (column) eigenvector and left (row) eigenvector, normalized to have unit trace: it is given by \eqref{projector}. 
\bl
\label{Lemma1}
Let $z_o$ be any points outside of the branching locus of $X:\CC \to\mathbb P^1$. Then 
\be
\sum_{a=1}^n \Cauchy_{\scr D-p_o}(p;z_o^{(a)} )\Cauchy_{\scr D-p_o}(z_o^{(a)} ; q)  = \Cauchy_{\scr D-p_o} (p;q) \frac {(z-x)\d z_o}{(z_o - x) (z_o- z) }
\ee
\el
{\bf Proof.}
The expression is a differential of $\d z_o$ on $\P^1$ with two simple poles at  $z_o = X(p) = x$ and $z_o = X(q) = z$.  The residue with respect to $z_o$ at $z_o=z$  of the left hand side is immediately computed  to be $\Cauchy_{\scr D - p_o}(p;q)$. The proof  follows. \QED

\bl
\label{lemmaproj}
Let $p=(x,y)\in \CC$; the spectral projector of $L(x)$ is given by the formula 
\be
\label{Piformula}
\Pi_{ab}(x,y) = (x-z_o)^2\frac {\Cauchy_{\scr D-p_o}(z_o^{(a)} ; p) \Cauchy_{\scr D-p_o}(p;z_o^{(b)} )}{\d x \d z_o} .
\ee
\el
{\bf Proof.} This is literally the outer product of the left and right eigenvectors. To verify that this is the projector we need to check that the trace is unity. To this end we observe that the trace of \eqref{Piformula} is the limit for $q\to p$ of the formula in  Lemma \ref{Lemma1}.  Because of the normalization $\res{p=q}\Cauchy_{\scr D-p_o}(p;q)=1$ we have 
\be
\Cauchy_{\scr D-p_o}(p;q) = \frac { \d x \d z}{(x-z)} + \mathcal O(1), \ \ \ q\to p.
\ee
From this the statement $\tr \, \Pi(x,y)\equiv 1$ follows. \QED

\paragraph {Proof of Corollary \ref{bidiffcoroll}.}
We need to compute $\tr \,\Pi(p)\Pi(q)$ with $p = (x,y), \ q= (z,w)$. 
Using the formula \eqref{Piformula} we have 
\bea
\tr \Pi(p)\Pi(q) = \sum_{a,b=1}^n
(x-z_o)^2\frac {\Cauchy_{\scr D-p_o}(z_o^{(a)} ; p) \Cauchy_{\scr D-p_o}(p;z_o^{(b)} )}{\d x \d z_o}
(z-z_o)^2\frac {\Cauchy_{\scr D-p_o}(z_o^{(b)} ; q) \Cauchy_{\scr D-p_o}(q;z_o^{(a)} )}{\d z \d z_o}.
\eea
The two sums are computed using Lemma \ref{Lemma1} and yield the desired result. \QED
\paragraph{Proof of Corollary \ref{polydiffcoroll}.}
This also follows immediately from  Lemma \ref{lemmaproj} and simple algebra. 
\subsection{$\scr D$--normalized first and second-kind differentials}
The Cauchy kernel \eqref{Cauchyvague} serves additional purposes: 
\begin{enumerate}
\item Consider the differentials (recall $p=(x,y), q = (z,w)$)
\be
\omega_j(p):= \res{q = d_j} \Cauchy_{\scr D-p_o}(p,q) \d z .
\ee
They are actually {\it holomorphic} differentials (the pole at $p=p_o$ disappears as can be seen by exchanging the order of residues); they replace the Legendre interpolation polynomials in the sense that they satisfy the interpolation problem 
\be
\omega_j(d_\ell) =0, \ \ \ \ell \neq j; \ \ \frac{ \omega_j(p)}{\d x} \bigg|_{p=d_j} =1. 
\ee
\item The differentials 
\be
\eta_{k}(p):= \res{q=p_o} \Cauchy(p,q) \frac {\d z}{(z-p_{ox})^{k+1}}
\ee
are the unique Abelian differentials of the second kind such that 
\be
\eta_k(p) = \le( \frac {1}{(x-p_{ox})^k}  + \mathcal O(1) \ri) \d x; \ \ \ \ 
\div(\eta_k) \geq - k(p_o) + \scr D.
\ee
\end{enumerate}

\subsection{Relationship with transcendental approaches}
\label{Szegosection}
The purpose of this section is to connect with other approaches to the inverse spectral problem, including \cite{BabelonBook, Bertola:EffectiveIMRN, BKS, Dubrovin}. 
 A way of presenting the approach, with the benefit of hindsight provided by \cite{Korotkin05}, is by the use of the Szeg\"o kernel:
 \be
 S_{\mathcal X} (p,q):= \frac {\Theta_{\mathcal X} (\int_q^p \vec\omega)}{\Theta_{\mathcal X} (0) \mathcal E(p,q)}
 \ee
 where $\mathcal E(p,q)$ is the Klein prime form (\cite{Faybook}, Ch. II), $\vec \omega = (\omega_1, \dots, \omega_g)$ is the vector of $a$--normalized abelian differentials  and $\bs \tau$ the corresponding matrix of $b$--periods
 \be
 \oint_{a_j} \omega_k = \delta_{jk}, \ \ \ \bs \tau_{jk} = \oint_{b_j} \omega_k, 
 \ee
  $\Theta_{\mathcal X}$ denotes the theta function with characteristic $\mathcal X = \bs \beta + \bs \tau \bs \alpha$, \ \ \ $\bs \alpha, \bs \beta\in \R^g$:
 \be
 \Theta_{\mathcal X}(\bs z) = \sum_{{\bf n } \in \Z^g} {\rm e}^{ i \pi \le({\bf n}+ \frac 1 2\bs \alpha\ri)^t \bs \tau \le({\bf n} + \frac 1 2 \bs \alpha \ri) - 2i \pi \le({\bf n} + \frac 1 2 \bs \alpha\ri)^t \le( \bs z  + \frac 12 \bs \beta\ri)}.
 \ee
 With respect to $p, q$ the Szeg\"o kernel is a section of a square-root of the canonical bundle (a ``half-differential'' or "semi-differential"). 
Another way of constructing the eigenvector matrix for the Lax matrix is to set 
\be
\Psi_{a,b} (x) = S_\mathcal X(x^{(b)}, z_o^{(a)}).
\ee
 This is clearly a different choice from the one in \eqref{PsiCauchy} but here we have the hint of the relationship:
  the $a$-th row  of $\Psi$ here, as before, consists of the evaluation on the different sheets of the semi-differential $\psi_a(p)  = S_{\mathcal X}(p, z_o^{(a)})$. The divisor  of its zeros, $ \scr D_a$, satisfies 
  \be
  \mathfrak A_{r_o}  \le(\scr  D_{a} - z_o^{(a)} \ri)+ \mathcal K_{r_o}= \mathcal X.
  \ee
  Here $\mathcal K_{r_o} $ is the vector of Riemann constants \cite{Faybook} and $\mathfrak A_{r_o} $ denotes the Abel map with base-point $r_o\in \CC$.  Note that while the Abel map and the vector of Riemann constants depend on the base-point , the combination above does not.  This happens because the divisor $\scr  D_{a} - z_o^{(a)}$ is of degree $g-1$ and $\mathcal K_{r_o} - \mathcal K_{\wt r _o} = (g-1)\mathfrak A(\wt r_o-r_o)$.
  
  In view of the equivalence \eqref{equivalence} $\scr D - p_o \equiv \scr D_{a} - z_o^{(a)}$ we conclude that 
\bea
S_\mathcal X(p,q) S_\mathcal X(q,p)  = \Cauchy_{\scr D-p_o}(p;q)\Cauchy_{\scr D-p_o}(q; p)\nn\\
 \mathcal X =  \mathfrak A_{r_o}  \le(\scr  D - p_o \ri)+ \mathcal K_{r_o}.
 \label{chardiv}
\eea
The interest of this relationship is that the Szeg\"o kernel on the left side is inherently a transcendental object, but the Cauchy kernel on the right side is an algebraic one,  given by Prop. \ref{propCauchy} and formula \eqref{Cauchyexp}.
This can be complemented with the formula of Fay (\cite{Faybook}, Ch. II, (39))
\be
S_\mathcal X(p,q) S_\mathcal X(q,p)  ={\bf B} (p,q) + \sum_{j,k=1}^g \omega_j(p) \omega_k(q) \frac {\pa^2 \ln \Theta_{\mathcal X}(0)}{ \pa z_j \pa  z_k}, 
\ee
with ${\bf B}(p,q)$ here denoting the $a$--normalized fundamental bi-differential.

%

\section{Examples}
\label{examples}
With the aid of a CAS one can easily produce examples of arbitrary size and genus, but the resulting numerical coefficients become rapidly too large to display on a paper. 
These examples have been constructed by choosing the general shape of the Newton polygon in advance, and then fixing the coefficients so that the curve passes through some chosen points with rational coordinates. 

\begin{example}
\rm
 The formula can be implemented with any computer algebra system.  Here is a nontrivial example;
 \bea
 L(x)  = \left[ \begin {array}{ccc} {\frac {20668\,{x}^{2}+38425\,x+16989}{768 (
x+3)}}&{\frac { \left( 5\,x+5 \right)  \left( 1708\,x+1329
 \right) }{384 ( x+3) }}&-{\frac { \left( 7\,x+7 \right)  \left( 5612
\,x+4161 \right) }{768 (x+ 3)}}\\ \noalign{\medskip}{\frac { \left( x
+1 \right)  \left( 253183\,x+379403 \right) }{15360 ( x+ 3) }}&{\frac 
{20923\,{x}^{2}+51570\,x+32183}{1536 ( x+3) }}&-{\frac { \left( 7\,x+7
 \right)  \left( 68747\,x+96391 \right) }{15360 (x+3)}}
\\ \noalign{\medskip}{\frac {5167\,x}{512}}+{\frac{5167}{512}}&{\frac 
{2135\,x}{256}}+{\frac{2135}{256}}&-{\frac {9821\,x}{512}}-{\frac{9053
}{512}}\end {array} \right]\\
E(x,y) =  \left( x+3 \right) {y}^{3}+ \left( -{\frac {2733\,{x}^{2}}{128}}-{
\frac {1609\,x}{192}}+{\frac{3829}{384}} \right) {y}^{2}+ \left( {x}^{
2}+x-\frac 1 2  \right) y+{x}^{2}-\frac 3 8\,x-\frac 5 8
\\ 
g = 2, \ \ \scr D -p_o=  \le(\frac 13 ,\frac 13\ri) + \le(\frac 23, \frac 43\ri) - (1,0),\nn\\ \ \ X^{-1}(\{-1\}) = \le\{
\le(-1, - \frac 1 2\ri), \le(-1, \frac 1 2 \ri), \le(-1, \frac 32 \ri)\ri\} .\nn\\
\hfill \blacktriangle
 \eea
\end{example}
\begin{example}
Here is an example in $g=2$
\bea
E(x,y) = &{y}^{3}x+{\frac {187\,{x}^{2}}{422}}+{\frac {597\,x}{422}}+{\frac{1007
}{1688}}+y \left( -{\frac {22639\,{x}^{2}}{6752}}-{\frac {64913\,x}{
20256}}+{\frac{2017}{5064}} \right) +{y}^{2} \left( -{\frac {6317\,x}{
2532}}-{\frac{2519}{2532}} \right) 
\nn\\
& \scr D -p_o= \le(-\frac 13,-2\ri) +\le(0,1\ri) - \le(-\frac 12,0\ri) 
\qquad
\wt{\scr D} -p_o= \le(\frac 13,-1\ri) +\le(2,\frac 14 \ri) - \le(-\frac 12,0\ri) 
\eea 
We have, corresponding to the two divisors;
\bea
 L(x) := A_1 x + A_0 + \frac {A_{-1}} x;  \ \ \wt L(x) := \wt A_1 x + \wt A_0 + \frac {\wt A_{-1} }x; 
\eea
\be
\wt L(x) = P_{\scr D, \wt {\scr D}}(x)\,  L (x)\,P_{\scr D, \wt {\scr D}}^{-1}(x).
\ee
\be
L(-1) = \wt L(-1) = {\rm diag} (-\frac 1 2, \frac 12  \frac 32).
\ee
{
\tiny
\bea
A_1 = \left[ \begin {array}{ccc} -{\frac{305013695}{547074048}}&{\frac{
36762211}{182358016}}&{\frac{146051771}{182358016}}
\\ \noalign{\medskip}{\frac{3587544275}{820611072}}&-{\frac{432393895}
{273537024}}&-{\frac{1717848095}{273537024}}\\ \noalign{\medskip}-{
\frac{2443006825}{1641222144}}&{\frac{294446885}{547074048}}&{\frac{
1169801485}{547074048}}\end {array} \right]
,\ 
A_0 =  \left[ \begin {array}{ccc} -{\frac{172225487}{273537024}}&{\frac{
23062403}{91179008}}&{\frac{62992043}{91179008}}\\ \noalign{\medskip}{
\frac{2476300115}{410305536}}&-{\frac{120515639}{136768512}}&-{\frac{
917432255}{136768512}}\\ \noalign{\medskip}-{\frac{2393754361}{
820611072}}&{\frac{100340389}{273537024}}&{\frac{1095694909}{273537024
}}\end {array} \right]
\nn\\
A_{-1} = \left[ \begin {array}{ccc} {\frac{234099745}{547074048}}&{\frac{
9362595}{182358016}}&-{\frac{20067685}{182358016}}
\\ \noalign{\medskip}{\frac{1365055955}{820611072}}&{\frac{18198035}{
91179008}}&-{\frac{117016415}{273537024}}\\ \noalign{\medskip}-{\frac{
2344501897}{1641222144}}&-{\frac{31255369}{182358016}}&{\frac{
200977261}{547074048}}\end {array} \right] 
\\
\wt A_1 = \left[ \begin {array}{ccc} -{\frac{99911711}{34192128}}&-{\frac{
15463713}{11397376}}&-{\frac{7817917}{11397376}}\\ \noalign{\medskip}{
\frac{271125907}{51288192}}&{\frac{13987727}{5698688}}&{\frac{21215129
}{17096064}}\\ \noalign{\medskip}{\frac{204290167}{102576384}}&{\frac{
10539587}{11397376}}&{\frac{15985349}{34192128}}\end {array} \right] 
\wt A_0 = \left[ \begin {array}{ccc} -{\frac{2862889}{534252}}&-{\frac{4417957}
{1424672}}&-{\frac{2142197}{712336}}\\ \noalign{\medskip}{\frac{
16008727}{1602756}}&{\frac{15363893}{2137008}}&{\frac{7348039}{1068504
}}\\ \noalign{\medskip}{\frac{2898439}{3205512}}&-{\frac{233527}{
4274016}}&{\frac{1419211}{2137008}}\end {array} \right] 
\\
\wt A_{-1} = \left[ \begin {array}{ccc} -{\frac{66217121}{34192128}}&-{\frac{
19879943}{11397376}}&-{\frac{26457235}{11397376}}\\ \noalign{\medskip}
{\frac{241153357}{51288192}}&{\frac{72399931}{17096064}}&{\frac{
96353495}{17096064}}\\ \noalign{\medskip}-{\frac{111540119}{102576384}
}&-{\frac{33486977}{34192128}}&-{\frac{44566165}{34192128}}
\end {array} \right] 
\eea
\be
P_{\scr D, \wt {\scr D}}=
\left[ \begin {array}{ccc} {\frac {191431\,{x}^{2}-147934\,x-15269}{
81024\,{x}^{2}-189056\,x+54016}}&{\frac {-9441\,{x}^{2}+58002\,x+67443
}{81024\,{x}^{2}-189056\,x+54016}}&{\frac {-14505\,{x}^{2}+88386\,x+
102891}{81024\,{x}^{2}-189056\,x+54016}}\\ \noalign{\medskip}{\frac {-
798767\,{x}^{2}-564994\,x+233773}{121536\,{x}^{2}-283584\,x+81024}}&{
\frac {-59757\,{x}^{2}-139942\,x+81863}{40512\,{x}^{2}-94528\,x+27008}
}&{\frac {-152597\,{x}^{2}-69046\,x+83551}{40512\,{x}^{2}-94528\,x+
27008}}\\ \noalign{\medskip}{\frac { \left( x+1 \right)  \left( 
1266313\,x-259691 \right) }{243072\,{x}^{2}-567168\,x+162048}}&{\frac 
{209979\,{x}^{2}+32826\,x-177153}{81024\,{x}^{2}-189056\,x+54016}}&{
\frac {400723\,{x}^{2}-139350\,x-215977}{81024\,{x}^{2}-189056\,x+
54016}}\end {array} \right] 
%
\ee

}\end{example}

\appendix
\section{Proof of Lemma \ref{lemmaOmega}}
\renewcommand{\theequation}{\Alph{section}.\arabic{equation}}

\label{exercise}
Inspection of \eqref{defOmega} shows that the only finite poles are for $(x,y) = (z,w), (p_{ox}, p_{oy})$. The remaining terms are polynomials, hence we need to verify that there are no poles when one of, or both $x,y$ tend to infinity along the curve. 

We will consider the case of assumption \eqref{demonic}, with respect to the variable $y$, the case with respect to $x$ being treated analogously after the swapping of roles:
\be
E(x,y) = a_n(x) y^n + \sum_{\ell=0}^{n-1} a_\ell(x) y^\ell, \ \ \ 
{\rm result}_x(a_n, a_{n-1}) \neq 0. 
\label{coeffs}
\ee
\paragraph{Case $x\to\infty, y\to \infty$.}
\bea
\Omega(p;q,p_o):= &\frac{ E(z, y )}{(x -z)(y-w)}  - \frac{ E(p_{o x}, y)}{(x - p_{ox})(y-p_{oy})}
 + y^{n-1} \le(\frac {a_n(z) - a_n(x)}{z-x}-  \frac {a_n(p_{ox}) - a_n(x)}{p_{ox}-x}  \ri)\nn\\
  &p=(x,y), \ \ q = (z,w).
  \label{defOmega3}
\eea
The first two terms are $\mathcal O(y^{n-1} x^{-1})$ while the last term is $\mathcal O(y^{n-1} x^{\deg a_n-1})$ and hence 
the most singular  behaviour of \eqref{thirdkind} is coming from the last term  
\be
\omega:= \Omega(p;q,p_o)\frac {\d x}{E_y(x,y)} \simeq  \frac{\cancel{y^{n-1}}}{n\cancel{y^{n-1}} \cancel{a_n(x) }}   \cancel{a_n(x)} \le(\frac 1{x-z} - \frac 1 {x-p_{ox}}\ri) =  \mathcal O(x^{-2}) \d x.
\ee 
Hence we have a regular differential at the common poles of $X, Y$. 
%

\paragraph{Case $y\to b\in \C$, $x\to \infty$.} Without loss of generality we assume $b=0$ (up to a translation in the $y$-variable). 
This case  can happen if and only if the Newton polygon has a side on the right of positive slope $\rho$ (facing ``south-east"). See Figure \ref{NewtonPolygon}.  Then $y =\mathcal O(x^{-\frac 1 \rho})$. 

Let $m_\ell = \deg_{x} a_\ell(x)$; let $\ell_1, \ell_2$ be the vertices bounding the side of the Newton polygon $\mathcal N$ corresponding to the branch $y(x)  = c x^{-\frac 1\rho}$. 
We write $\omega = -\Omega(p;q,p_o) \frac {\d x}{E_y(x,y)}$.
Then 
$$
E_y(x,y) =\mathcal O (x^{m_{\ell_1} - \frac {\ell_1-1}\rho}), \ \ m_{\ell_1} - \frac {\ell_1-1}\rho \geq 0.
$$
As $x\to\infty$ and $ y= \mathcal O(x^{-\frac 1 \rho}) \to 0$, the first two terms in $\Omega$ \eqref{defOmega3} are $\mathcal O(x^{-1})$ and hence of no concern. The last term contains the expression $y^{n-1} a_n(x) \le(\frac 1{x-z} - \frac 1 {x-p_{ox}} \ri)$ which  is of order $\mathcal O(y^{n-1} x^{m_n-2}) = \mathcal O(x^{ m_n - \frac {n}\rho  - 2 + \frac 1{\rho}})$ so that in total
\be
\omega = \frac{\Omega \d x}{E_y(x,y)} = \mathcal O\le( \frac{x^{ m_n - \frac {n-1}\rho  - 2 }}{x^{m_1-\frac {\ell_1-1}{\rho} }}\ri) \d x.
\ee
Thus the exponent of $x$ in the above expression is $(m_n-m_1) - \frac{ n-\ell_1}\rho - 2$: since the point $(m_n, n)$ in the $\mu-\nu$ plane is above the line
$
\scr L:=\{\frac{(\nu-\ell_1) }\rho= \mu-m_1\} \subset \C_\mu\times \C_\nu
$, then $(m_n-m_1) - \frac{ n-\ell_1}\rho\leq 0$ and thus we have 
\be
\omega = \frac{\Omega \d x}{E_y(x,y)} = \mathcal O\le(x^{-2}\ri) \d x
\ee
and hence analytic as $x\to\infty$.

\paragraph{Case $y\to \infty, \ x\to b$.}
Again, we assume $b=0$ up to an $x$-translation. Thus $a_n(x) = x^\mu(C + \mathcal O(x))$ and $a_{n-1}(x) = \wt C(1 + \mathcal O(x))$, with $C, \wt C\neq 0$, as $x\to 0$; moreover $y = \mathcal O(x^{-\mu})$.
The expression \eqref{defOmega} for $\Omega$ is then easily verified to be of order $\mathcal O(y^{n-2}) = x^{(2-n) \mu}$.
Indeed, simplifying the expression we obtain 
\bea
\Omega(p,q;p_o)= y^{n-1} a_n(x)\le(\frac 1{ x-z} - \frac 1{x-p_{ox}} \ri) + \mathcal O(y^{n-2}).
\eea
Since $a_n(x) = \mathcal O(x^\mu)$ also the first term is $\mathcal O(y^{n-2})$. 
 But here also $\frac {\d x}{E_y(x,y)} = \mathcal O\le( y^{2-n} \d x\ri) = \mathcal O\le( x^{(n-2)\mu} \d x\ri) $ so that the expression is, once more, a differential regular at $x=\infty$.

\paragraph{Acknowledgments.}
The author completed the work during his tenure as Royal Society Wolfson Visiting
Fellow (RSWVF/R2/242024) at the School of Mathematics in Bristol University. The work was supported in part by the Natural Sciences and Engineering Research Council of Canada (NSERC) grant RGPIN-2023-04747.

\paragraph{ Conflict of Interest declaration.}
The authors have no financial or proprietary interests in any material discussed in this article.

\paragraph{ Data availability.}
No data was produced or analyzed in this article.

\end{document}